\documentclass[aps,prl,twocolumn,superscriptaddress,showpacs,preprintnumbers,amsmath,amssymb,floatfix]{revtex4}

\usepackage{graphicx}
\usepackage{dcolumn}
\usepackage{bm}
\usepackage{url} 
\usepackage{amsmath} 
\usepackage{amssymb} 

\begin{document}

\title{\quad\\[1.0cm] 
Search for $CP$ Violation in the Decays $D^{+}_{(s)}\rightarrow K^0_S\pi^+$ and
$D^{+}_{(s)}\rightarrow K^0_S K^+$
}

\affiliation{Budker Institute of Nuclear Physics, Novosibirsk}
\affiliation{Faculty of Mathematics and Physics, Charles University, Prague}

\affiliation{University of Cincinnati, Cincinnati, Ohio 45221}

\affiliation{Justus-Liebig-Universit\"at Gie\ss{}en, Gie\ss{}en}
\affiliation{The Graduate University for Advanced Studies, Hayama}

\affiliation{Hanyang University, Seoul}
\affiliation{University of Hawaii, Honolulu, Hawaii 96822}
\affiliation{High Energy Accelerator Research Organization (KEK), Tsukuba}

\affiliation{Institute of High Energy Physics, Chinese Academy of Sciences, Beijing}
\affiliation{Institute of High Energy Physics, Vienna}
\affiliation{Institute of High Energy Physics, Protvino}

\affiliation{Institute for Theoretical and Experimental Physics, Moscow}
\affiliation{J. Stefan Institute, Ljubljana}
\affiliation{Kanagawa University, Yokohama}
\affiliation{Institut f\"ur Experimentelle Kernphysik, Karlsruhe Institut f\"ur Technologie, Karlsruhe}
\affiliation{Korea University, Seoul}

\affiliation{Kyungpook National University, Taegu}
\affiliation{\'Ecole Polytechnique F\'ed\'erale de Lausanne (EPFL), Lausanne}
\affiliation{Faculty of Mathematics and Physics, University of Ljubljana, Ljubljana}
\affiliation{University of Maribor, Maribor}
\affiliation{Max-Planck-Institut f\"ur Physik, M\"unchen}
\affiliation{University of Melbourne, School of Physics, Victoria 3010}
\affiliation{Nagoya University, Nagoya}

\affiliation{Nara Women's University, Nara}
\affiliation{National Central University, Chung-li}
\affiliation{National United University, Miao Li}
\affiliation{Department of Physics, National Taiwan University, Taipei}
\affiliation{H. Niewodniczanski Institute of Nuclear Physics, Krakow}
\affiliation{Nippon Dental University, Niigata}
\affiliation{Niigata University, Niigata}
\affiliation{University of Nova Gorica, Nova Gorica}
\affiliation{Novosibirsk State University, Novosibirsk}
\affiliation{Osaka City University, Osaka}

\affiliation{Panjab University, Chandigarh}

\affiliation{RIKEN BNL Research Center, Upton, New York 11973}

\affiliation{University of Science and Technology of China, Hefei}
\affiliation{Seoul National University, Seoul}

\affiliation{Sungkyunkwan University, Suwon}
\affiliation{School of Physics, University of Sydney, NSW 2006}
\affiliation{Tata Institute of Fundamental Research, Mumbai}
\affiliation{Excellence Cluster Universe, Technische Universit\"at M\"unchen, Garching}
\affiliation{Toho University, Funabashi}
\affiliation{Tohoku Gakuin University, Tagajo}
\affiliation{Tohoku University, Sendai}
\affiliation{Department of Physics, University of Tokyo, Tokyo}

\affiliation{Tokyo Metropolitan University, Tokyo}
\affiliation{Tokyo University of Agriculture and Technology, Tokyo}

\affiliation{IPNAS, Virginia Polytechnic Institute and State University, Blacksburg, Virginia 24061}
\affiliation{Yonsei University, Seoul}
  \author{B.~R.~Ko}\affiliation{Korea University, Seoul} 
  \author{E.~Won}\affiliation{Korea University, Seoul} 

  \author{H.~Aihara}\affiliation{Department of Physics, University of Tokyo, Tokyo} 
  \author{V.~Aulchenko}\affiliation{Budker Institute of Nuclear Physics, Novosibirsk}\affiliation{Novosibirsk State University, Novosibirsk} 
  \author{T.~Aushev}\affiliation{\'Ecole Polytechnique F\'ed\'erale de Lausanne (EPFL), Lausanne}\affiliation{Institute for Theoretical and Experimental Physics, Moscow} 
  \author{A.~M.~Bakich}\affiliation{School of Physics, University of Sydney, NSW 2006} 
  \author{V.~Balagura}\affiliation{Institute for Theoretical and Experimental Physics, Moscow} 

  \author{E.~Barberio}\affiliation{University of Melbourne, School of Physics, Victoria 3010} 
  \author{K.~Belous}\affiliation{Institute of High Energy Physics, Protvino} 
  \author{V.~Bhardwaj}\affiliation{Panjab University, Chandigarh} 
% ...
\collaboration{The Belle Collaboration}
\noaffiliation
 
\begin{abstract}
We have searched for $CP$ violation in the charmed meson decays
$D^{+}_{(s)}\rightarrow K^0_S\pi^+$ and $D^{+}_{(s)}\rightarrow K^0_S K^+$
using 673 fb$^{-1}$ of data collected with the Belle detector at the KEKB
asymmetric-energy $e^+e^-$ collider. No evidence for $CP$ violation is
observed. We report the most sensitive $CP$ asymmetry measurements to date for
these decays: $A_{CP}^{D^+\rightarrow K^0_S\pi^+}=(-0.71\pm0.19\pm0.20)\%$,
$A_{CP}^{D^+_s\rightarrow K^0_S\pi^+}=(+5.45\pm2.50\pm0.33)\%$,
$A_{CP}^{D^+\rightarrow K^0_S K^+}=(-0.16\pm0.58\pm0.25)\%$, and
$A_{CP}^{D^+_s\rightarrow K^0_S K^+}=(+0.12\pm0.36\pm0.22)\%$, where the first
uncertainties are statistical and the second are systematic.
\end{abstract}

\pacs{11.30.Er, 13.25.Ft, 14.40.Lb}

\maketitle

{\renewcommand{\thefootnote}{\fnsymbol{footnote}}}
\setcounter{footnote}{0}
Violation of the combined Charge-conjugation and Parity symmetries ($CP$) in
the standard model (SM) is produced by a non-vanishing phase in the
Cabibbo-Kobayashi-Maskawa flavor-mixing matrix~\cite{CKM}. For charged meson
decays this may be observed as a non-zero $CP$ asymmetry, defined as
\begin{equation}
  A_{CP}=\frac
  {\Gamma(X^+\rightarrow f^+)-\Gamma(X^-\rightarrow f^-)}
  {\Gamma(X^+\rightarrow f^+)+\Gamma(X^-\rightarrow f^-)}
  \label{EQ:ACP}
\end{equation}
where $\Gamma$ is the partial decay width,  $X$ denotes a charged meson, and
$f$ is a final state.

In the SM, the charmed particle processes for which a significant non-vanishing
$CP$ violation is expected are singly Cabibbo-suppressed (SCS) decays in which
there is both interference between two different decay amplitudes and a strong
phase shift from final state interactions. In the SM, $CP$ violation in SCS
charmed meson decays is predicted to occur at the level of $\mathcal{O}(0.1)$\%
or lower~\cite{SMCP}. The SM also predicts a $CP$ asymmetry in the final states
containing a neutral kaon that is produced via $K^0-\bar{K}^0$ mixing even if
no $CP$ violating phase exists in the charm decay amplitudes. The expected
magnitude for this type of asymmetry is $(0.332\pm0.006)$\% from $K^0_L$
semileptonic decay~\cite{PDG2008}. Searches for $CP$ violation in charmed
mesons are complementary to those in $B$ and $K$ mesons, since the former tests
the $CP$ violating couplings of the $up$-type quarks while the latter those of
the $down$-type quarks.

In this Letter we report results from searches for $CP$ violation in the
$D^+_s\rightarrow K^0_S\pi^+$, $D^+\rightarrow K^0_S K^+$, $D^+\rightarrow
K^0_S\pi^+$, and $D^+_s\rightarrow K^0_S K^+$ decay processes~\cite{CC}. The
former two channels are SCS decays and the latter two are  mixtures of
Cabibbo-favored (CF) and doubly Cabibbo-suppressed (DCS) decays, where SM $CP$
violations described above are expected. Observing $A_{CP}$ values of
$\mathcal{O}(1)$\% or larger in the decays considered in this Letter would
represent strong evidence for processes involving physics beyond the
SM~\cite{BIGI}.

The data used in this analysis were recorded at or near the $\Upsilon(4S)$
resonance with the Belle detector~\cite{BELLE} at the $e^+e^-$
asymmetric-energy collider KEKB~\cite{KEKB}. The sample corresponds to an
integrated luminosity of 673 fb$^{-1}$.

\begin{figure}[htbp]
\mbox{
  \includegraphics[height=0.25\textwidth,width=0.47\textwidth]{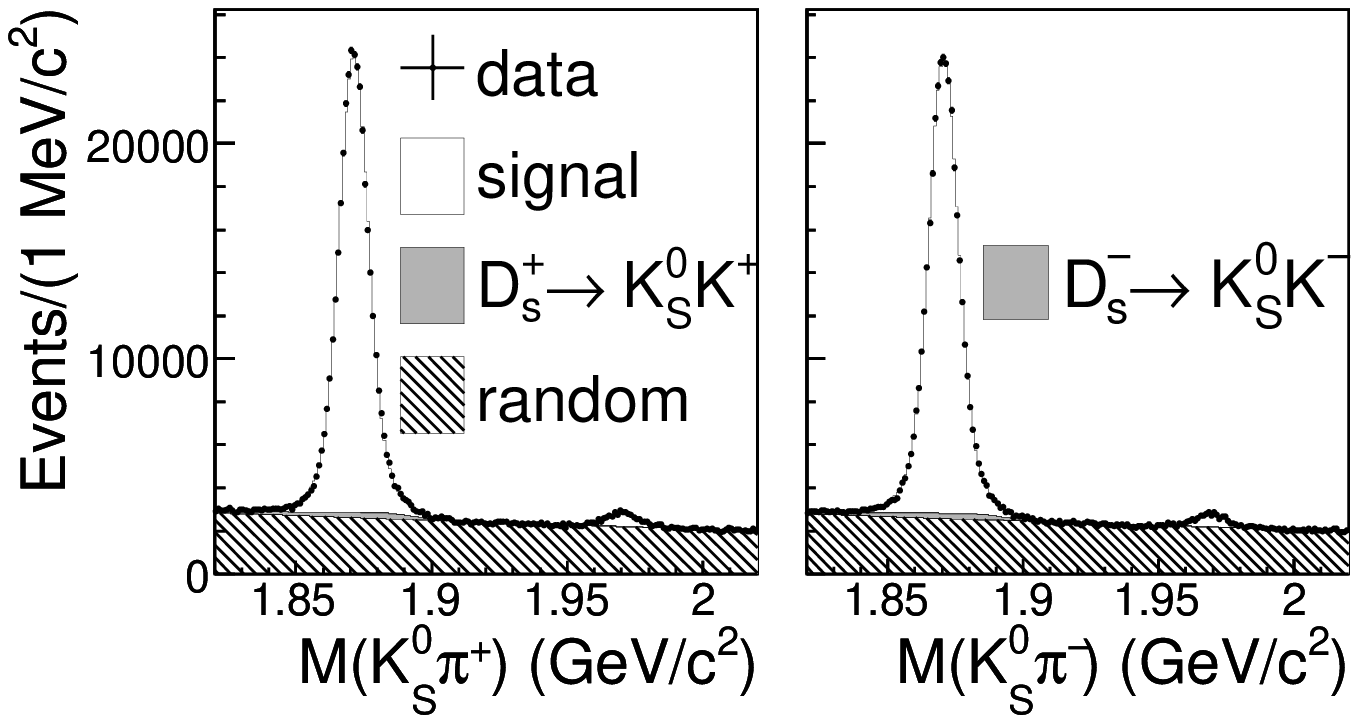}
}
\mbox{
  \includegraphics[height=0.25\textwidth,width=0.47\textwidth]{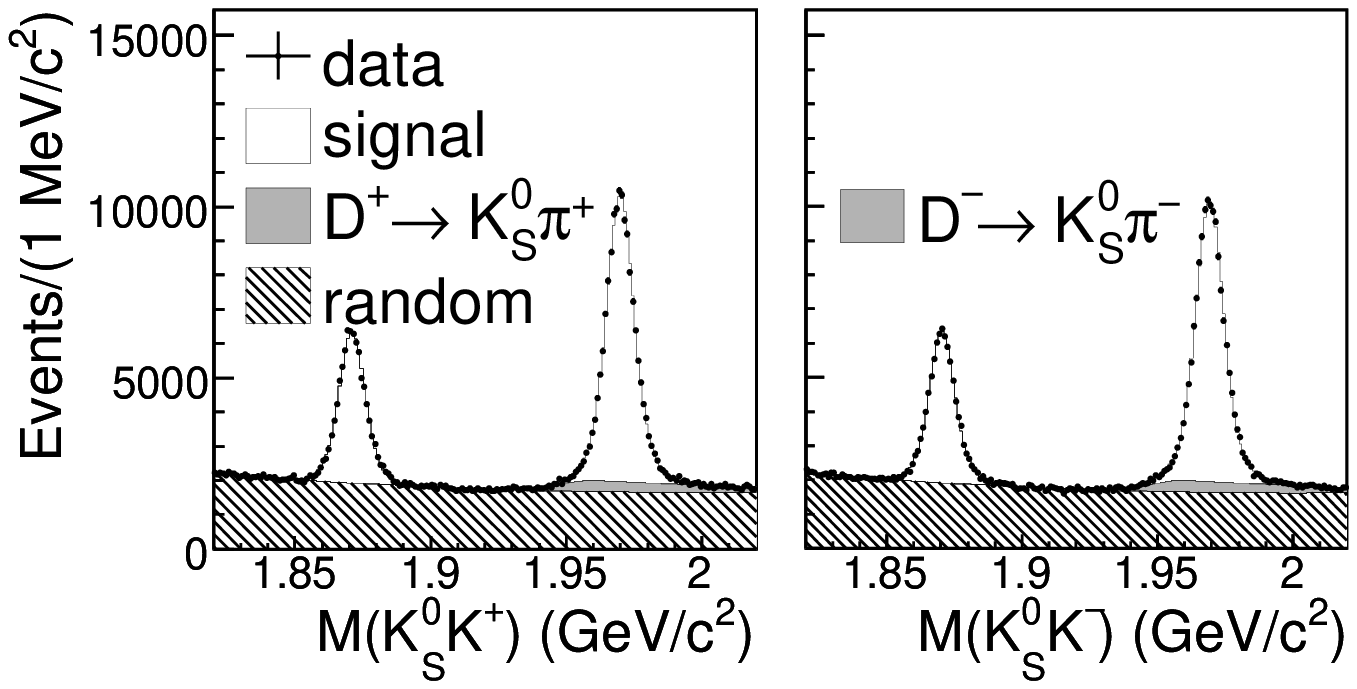}
}
\caption{Invariant mass distributions for the $K^0_S\pi^{\pm}$ and $K^0_S
  K^{\pm}$ final states. Points with error bars (note the small size of them
  due to the large sample) show the data and the histograms show the results of
  the parameterizations of the data. Signal, peaking background, and random
  combinatorial background components are also shown.}
\label{FIG:MKSX}
\end{figure}

$D^+$ and $D^+_s$ candidates are reconstructed using the same requirements used
in the measurement of the branching ratios for these  same decays reported in
Ref.~\cite{BRKS}. Figure~\ref{FIG:MKSX} shows the reconstructed
$K^0_S\pi^{\pm}$ and $K^0_S K^{\pm}$ invariant mass distributions. All signals
are parameterized as a sum of two Gaussian distributions except for
$D^+_s\rightarrow K^0_S\pi^+$ in which a single Gaussian is used for the signal
parameterization. The parameterizations of the random combinatorial background
and the peaking background due to $K/\pi$ misidentification are described in
detail in Ref.~\cite{BRKS}. 

We determine the quantities  $A^{X^+\rightarrow K^0_S h^+}_{CP}$ defined in
Eq.~(\ref{EQ:ACP}) by measuring the signal yield asymmetry
\begin{equation}
  A^{X^+\rightarrow K^0_S h^+}_{\rm rec}~=~\frac
  {N_{\rm rec}^{X^+\rightarrow K^0_S h^+}-N_{\rm rec}^{X^-\rightarrow K^0_S h^-}}
  {N_{\rm rec}^{X^+\rightarrow K^0_S h^+}+N_{\rm rec}^{X^-\rightarrow K^0_S h^-}},
  \label{EQ:ARECON}
\end{equation}
where $N_{\rm rec}$ is the number of reconstructed decays and $h$ is a charged
hadron. The measured asymmetry in Eq.~(\ref{EQ:ARECON}) includes two
contributions other than $A_{CP}$. One is the forward-backward asymmetry
($A_{FB}$) due to $\gamma^{*}-Z^0$ interference in $e^+e^-\rightarrow c\bar{c}$
and the other is a detection efficiency asymmetry between positively and
negatively charged tracks ($A^{h^+}_{\epsilon}$). Since $K^0_S$ mesons are
reconstructed from a $\pi^+\pi^-$ pair, there is no detection asymmetry other
than $A^{h^+}_{\epsilon}$. Eq.~(\ref{EQ:ARECON}) can therefore be expressed as
\begin{equation}
  A^{X^+\rightarrow K^0_S h^+}_{\rm rec}~=~A^{X^+\rightarrow K^0_S h^+}_{CP}~+~A^{X^+}_{FB}~+~A^{h^+}_{\epsilon}.
  \label{EQ:ARECONII}
\end{equation}
To correct for the asymmetries other than $A_{CP}$, we use reconstructed
samples of $D^+_s\rightarrow\phi\pi^+$ and $D^0\rightarrow K^-\pi^+$ decays and
assume that $A_{CP}$ in CF decays is negligibly small at the current
experimental sensitivity and that $A_{FB}$ is the same for all charmed
mesons. We reconstruct $\phi$ mesons via their $K^+K^-$ decay channel for
$D^+_s\rightarrow\phi\pi^+$, requiring the $K^+K^-$ invariant mass to be
between 1.01 and 1.03 GeV/$c^2$.

\begin{figure}[htbp]
  \includegraphics[height=0.5\textwidth,width=0.50\textwidth]{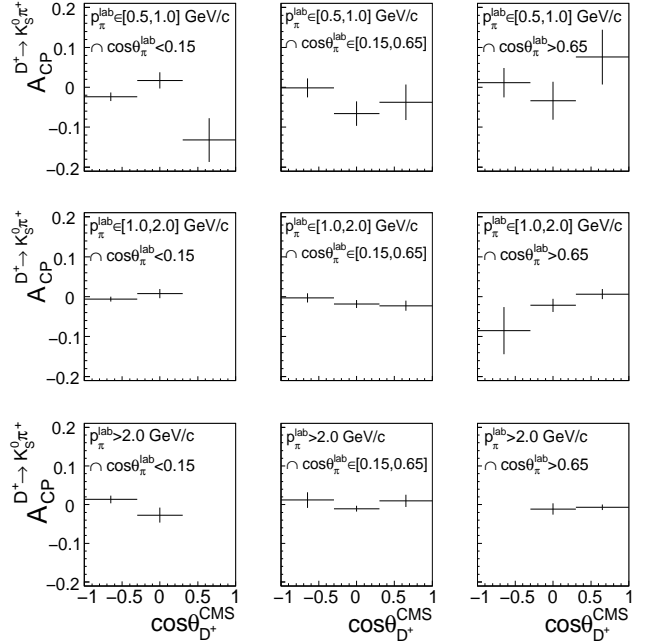}  
\caption{Measured $A_{CP}$ values for $D^+\rightarrow K^0_S\pi^+$ in bins of
  ($p^{\rm lab}_{\pi}$, $\cos\theta^{\rm lab}_{\pi}$, $\cos\theta^{\rm
    CMS}_{D^+}$). Empty bins where no entries are plotted have no statistics.}
\label{FIG:ACPKSPI}
\end{figure}

The measured asymmetry for $D^+_s\rightarrow\phi\pi^+$ is the sum of
$A^{D^+_s}_{FB}$ and $A^{\pi^+}_{\epsilon}$. Hence one can extract the $A_{CP}$
value for the $K^0_S\pi^+$ final states by subtracting the measured asymmetry for
$D^+_s\rightarrow\phi\pi^+$ from that for $D^+_{(s)}\rightarrow K^0_S\pi^+$. The
subtraction is performed in bins of $\pi^+$ momentum, $p^{\rm lab}_{\pi}$, and
polar angle in the laboratory system, $\cos\theta^{\rm lab}_{\pi}$  (because
$A^{h^+}_{\epsilon}$ depends on these two variables while it is uniform in
azimuthal angle), and the charmed meson's polar angle in the center-of-mass
system, $\cos\theta^{\rm CMS}_{D^+_{(s)}}$ (since $\cos\theta^{\rm
  CMS}_{D^+_{(s)}}$ is correlated with $\cos\theta^{\rm lab}_{\pi}$ and
$A_{FB}^{D^+_{(s)}}$ depends on it). The choice of the three-dimensional (3-D)
binning is selected in order to avoid large statistical fluctuations in each
bin. Figure~\ref{FIG:ACPKSPI} shows the $A_{CP}$ map of $D^+\rightarrow
K^0_S\pi^+$ in bins of ($p^{\rm lab}_{\pi}$, $\cos\theta^{\rm lab}_{\pi}$,
$\cos\theta^{\rm CMS}_{D^+_{(s)}}$). Calculating a weighted average of the
$A_{CP}$ values over the 3-D bins, we obtain $A_{CP}^{D^+\rightarrow
  K^0_S\pi^+}=(-0.71\pm0.26)\%$ where the uncertainty originates from the
finite size of the $D^+\rightarrow K^0_S\pi^+$ (0.19\%) and
$D^+_s\rightarrow\phi\pi^+$ (0.18\%) samples. The $\chi^2$/d.o.f over the 3-D
bins is found to be 31.4/24 which corresponds to 14\% probability.

The statistical precision of the $D^+_s\rightarrow K^0_S\pi^+$ sample is too
low to allow for a 3-D correction to $A_{\rm rec}^{D^+_s\rightarrow
  K^0_S\pi^+}$. For this mode we correct for asymmetries other than $A_{CP}$
with an inclusive correction obtained by subtracting $A_{\rm
  rec}^{D^+\rightarrow K^0_S\pi^+}$ from $A_{CP}^{D^+\rightarrow K^0_S\pi^+}$
after integrating over the entire ($p^{\rm lab}_{\pi}$, $\cos\theta^{\rm
  lab}_{\pi}$, $\cos\theta^{\rm CMS}_{D^+}$) space. The inclusive correction is
$(-0.34\pm0.18)\%$ where the uncertainty is entirely due to the statistical
uncertainty of the $D^+_s\rightarrow\phi\pi^+$
sample. The value of $A_{CP}^{D^+_s\rightarrow K^0_S\pi^+}$ is measured to be
$(+5.45\pm2.50)\%$, where the uncertainty is statistical only.

The dominant source of systematic uncertainty in the
$A_{CP}^{D^+_{(s)}\rightarrow K^0_S\pi^+}$ measurement is the uncertainty in
the $A_{\rm rec}^{D^+_s\rightarrow\phi\pi^+}$ determination, which originates from
the following sources: the statistical uncertainty of the selected
$D^+_s\rightarrow\phi\pi^+$ sample (0.18\%); the choice of the $M(K^+K^-)$
interval (0.03\%); and the choice of binning for the 3-D map of $A_{\rm
  rec}^{D^+_s\rightarrow\phi\pi^+}$ (0.03\%). Another source is the choice of
fitting parameters for the invariant mass distributions: binnings, mass
windows, and background parameterizations together, contribute uncertainties of
0.04\% to $A_{CP}^{D^+\rightarrow K^0_S\pi^+}$ and 0.27\% to
$A_{CP}^{D^+_s\rightarrow K^0_S\pi^+}$, where the larger uncertainty on
$A_{CP}^{D^+_s\rightarrow K^0_S\pi^+}$ is inherited from the low statistics of
$D^+_s\rightarrow K^0_S\pi^+$. We also consider possible effects due to the
differences in interactions of $K^0$ and $\bar{K}^0$ mesons with the material
of the detector. $K^0$ and $\bar{K}^0$ mesons considered in this Letter are
produced through the weak interaction and interact with the material near the
interaction point until they decay into $\pi^+\pi^-$ pairs. This produces a
non-vanishing asymmetry originating from the different strong interactions of
$K^0$ and $\bar{K}^0$ mesons with nucleons. Assuming that the differences
between $K^0$ and $\bar{K}^0$ interactions with nucleons are the same as those
for $K^+$ and $K^-$ interactions, we calculate the probability of $K^0$ and
$\bar{K}^0$-nucleons interactions using the known $K^+$ and $K^-$ cross
sections~\cite{PDG2008} and take into account the time evolution of neutral
kaons. We consider the beam pipe~\cite{BELLE} and the silicon vertex
detector~\cite{SVD2} in our estimates of the $K^0/\bar{K}^0$-material
effects. The uncertainty in the $CP$ asymmetry due to $K^0/\bar{K}^0$-material
effects is found to be 0.06\%. A summary of systematic uncertainties in
$A_{CP}^{D^+_{(s)}\rightarrow K^0_S\pi^+}$ is given in
Table~\ref{TABLE:SYSTOTAL}. By combining all systematic uncertainties in
quadrature, we obtain $A_{CP}^{D^+\rightarrow
  K^0_S\pi^+}=(-0.71\pm0.19\pm0.20)\%$ and $A_{CP}^{D^+_s\rightarrow
  K^0_S\pi^+}=(+5.45\pm2.50\pm0.33)\%$, where the first uncertainties are
statistical and the second are systematic.

The method for the measurement of $A_{CP}$ in the $K^0_S K^+$ final states is
different from that for the $K^0_S\pi^+$ final states. The $A^{D^+_{(s)}}_{FB}$
and $A^{\pi^+}_{\epsilon}$ components in $A^{D^+_{(s)}\rightarrow
  K^0_S\pi^+}_{\rm rec}$ are directly obtained from the
$D^+_s\rightarrow\phi\pi^+$ sample, but there is no corresponding large
statistics decay mode that can be used to directly measure the
$A^{D^+_{(s)}}_{FB}$ and $A^{K^+}_{\epsilon}$ components in
$A^{D^+_{(s)}\rightarrow K^0_S K^+}_{\rm rec}$. Thus, to correct the
reconstructed asymmetry in the $K^0_S K^+$ final states, we use samples of
$D^0\rightarrow K^-\pi^+$ as well as $D^+_s\rightarrow\phi\pi^+$ decays.

\begin{figure}[htbp]
\includegraphics[height=0.40\textwidth,width=0.50\textwidth]{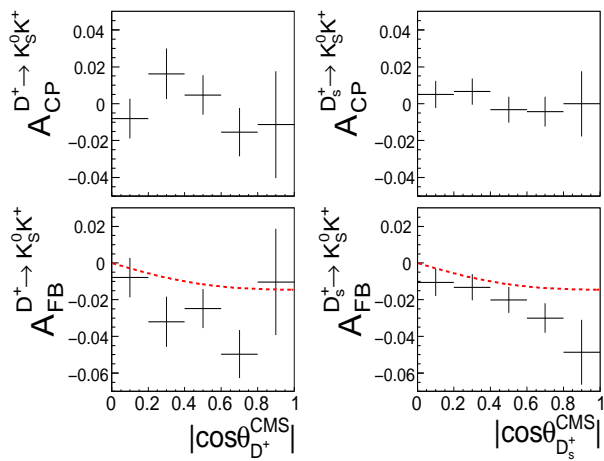}
\caption{Measured $A_{CP}$ and $A_{FB}$ values for $D^+_{(s)}\rightarrow K^0_S K^+$ as
  a function of $|\cos\theta^{\rm CMS}_{D^+_{(s)}}|$. The dashed curves show
  the leading-order prediction for $A^{c\bar{c}}_{FB}$.}
\label{FIG:ACPKSK}
\end{figure}

The measured asymmetry for $D^0\rightarrow K^-\pi^+$ is a sum of
$A^{D^0}_{FB}$, $A^{K^-}_{\epsilon}$, and $A^{\pi^+}_{\epsilon}$. Thus, we can
extract $A^{K^-}_{\epsilon}$ by subtracting the measured asymmetry for
$D^+_s\rightarrow\phi\pi^+$ from that for $D^0\rightarrow K^-\pi^+$. An
$A^{K^-}_{\epsilon}$ correction map is obtained as follows; $N^{D^0\rightarrow
  K^-\pi^+}_{\rm rec}$ and $N^{\bar{D}^0\rightarrow K^+\pi^-}_{\rm rec}$
are corrected according to the reconstructed asymmetry for
$D^+_s\rightarrow\phi\pi^+$ in bins of ($p^{\rm lab}_{\pi}$, $\cos\theta^{\rm
  lab}_{\pi}$, $\cos\theta^{\rm CMS}_{D}$). Subsequently, corrected
$N^{D^0\rightarrow K^-\pi^+}_{\rm rec}$ and $N^{\bar{D}^0\rightarrow
  K^+\pi^-}_{\rm rec}$ values are determined in bins of $K^{\mp}$ momentum and
polar angle in the laboratory frame, ($p^{\rm lab}_{K^{\mp}}$, $\cos\theta^{\rm
  lab}_{K^{\mp}}$). From the corrected values of $N^{D^0\rightarrow K^-\pi^+}_{\rm rec}$
and $N^{\bar{D}^0\rightarrow K^+\pi^-}_{\rm rec}$ in bins of ($p^{\rm
  lab}_{K^{\mp}}$, $\cos\theta^{\rm lab}_{K^{\mp}}$), we obtain an
$A^{K^-}_{\epsilon}$ map that is used to correct $A^{K^+}_{\epsilon}$~\cite{DEFINEASYM} for
the $K^0_S K^+$ final states. By subtracting $A^{K^+}_{\epsilon}$ from the
reconstructed asymmetry of $D^+_{(s)}\rightarrow K^0_S K^+$, we obtain the
corrected reconstruction asymmetry $A^{K^+}_{\epsilon}$ for
$D^+_{(s)}\rightarrow K^0_S K^+$:
\begin{equation} 
  \begin{split}
    & A^{D^+_{(s)}\rightarrow K^0_S K^+_{\rm corr}}_{\rm
      rec}~=~A^{D^+_{(s)}\rightarrow K^0_S K^+}_{\rm rec}~-~A^{K^+}_{\epsilon} \\
    &~~~~~~~~~~~~~~~~~~~~=~A^{D^+_{(s)}}_{FB}~+~A^{D^+_{(s)}\rightarrow K^0_S K^+}_{CP}.
    \end{split}
  \label{EQ:AKSK}
\end{equation}
As shown in Eq.~(\ref{EQ:AKSK}), $A^{D^+_{(s)}\rightarrow K^0_S K^+_{\rm
    corr}}_{\rm rec}$ includes not only an $A_{CP}$ component but also an
    $A_{FB}$ component. Since $A_{CP}$ is independent of all kinematic
    variables, while $A_{FB}$ is an odd function of $\cos\theta^{\rm
    CMS}_{D^+_{(s)}}$, and thus vanishes when integrated over it, we measure
    $A^{D^+_{(s)}\rightarrow K^0_S K^+_{\rm corr}}_{\rm rec}$ as a function of
    $\cos\theta^{\rm CMS}_{D^+_{(s)}}$. The $A_{CP}$ component in
    Eq.~(\ref{EQ:AKSK}) is then extracted according to Eq.~(\ref{EQ:ACPFB1}),
    using the above symmetry properties. We also extract the $A_{FB}$ component
    in Eq.~(\ref{EQ:AKSK}) using Eq.~(\ref{EQ:ACPFB2}).
\begin{subequations}
  \begin{equation} 
    \begin{split}
      & A^{D^+_{(s)}\rightarrow K^0_S K^+}_{CP}~=~[A^{D^+_{(s)}\rightarrow K^0_S K^+_{\rm corr}}_{\rm rec}(\cos\theta^{\rm CMS}_{D^+_{(s)}}) \\
	&~~~~~~~~~~~~~~~~~~+~A^{D^+_{(s)}\rightarrow K^0_S K^+_{\rm corr}}_{\rm rec}(-\cos\theta^{\rm CMS}_{D^+_{(s)}})]/2, \\%
    \end{split}
    \label{EQ:ACPFB1}
  \end{equation}
  \begin{equation}
    \begin{split} 
      & A^{D^+_{(s)}\rightarrow K^0_S K^+}_{FB}~=~[A^{D^+_{(s)}\rightarrow K^0_S K^+_{\rm corr}}_{\rm rec}(\cos\theta^{\rm CMS}_{D^+_{(s)}}) \\
	&~~~~~~~~~~~~~~~~~~-~A^{D^+_{(s)}\rightarrow K^0_S K^+_{\rm corr}}_{\rm rec}(-\cos\theta^{\rm CMS}_{D^+_{(s)}})]/2.
    \end{split}
    \label{EQ:ACPFB2}
  \end{equation}
\end{subequations}
Figure~\ref{FIG:ACPKSK} shows $A^{D^+_{(s)}\rightarrow K^0_S K^+}_{CP}$ and
  $A^{D^+_{(s)}\rightarrow K^0_S K^+}_{FB}$ as a function of $|\cos\theta^{\rm
  CMS}_{D^+_{(s)}}|$. Calculating a weighted average over the $|\cos\theta^{\rm
  CMS}_{D^+_{(s)}}|$ bins, we obtain $A^{D^+\rightarrow K^0_S
  K^+}_{CP}=(-0.16\pm0.58)\%$ and $A^{D^+_{s}\rightarrow K^0_S
  K^+}_{CP}=(+0.12\pm0.36)\%$ where the uncertainties are statistical only. The
  observed $A_{FB}$ values decrease with $\cos\theta^{\rm CMS}_{D^+_{(s)}}$ as
  expected from the leading-order prediction~\cite{AFBCC}. The observed
  deviations from the prediction are expected due to higher order corrections,
  and are in agreement with the measured asymmetries in the $K^+K^-$ and
  $\pi^+\pi^-$ final states~\cite{ACPD0}.
\begin{table*}[htbp]
\caption{\label{TABLE:SYSTOTAL} Summary of systematic uncertainties. $\sigma
  A_{CP}$ is the systematic uncertainty in $A_{CP}$.}
\begin{ruledtabular}
\begin{tabular}{cccccc} 
Source&                               &$\sigma A_{CP}^{D^+\rightarrow K^0_S\pi^+}$(\%)
                                      &$\sigma A_{CP}^{D^+_s\rightarrow K^0_S\pi^+}$(\%)
                                      &$\sigma A_{CP}^{D^+\rightarrow K^0_S K^+}$(\%)
                                      &$\sigma A_{CP}^{D^+_s\rightarrow K^0_S K^+}$(\%)\\ \hline
                                      &$D^+_s\rightarrow\phi\pi^+$ statistics &0.18&0.18&-&-\\ 
$A^{D^+_s\rightarrow\phi\pi^+}_{\rm rec}$ &$A^{D^+_s\rightarrow\phi\pi^+}_{\rm rec}$ binning &0.03&0.03&-&-\\
                                      &$M(K^+K^-)$ window &0.03&0.03&-&-\\ \hline
                                         &$D^+_s\rightarrow\phi\pi^+$ statistics&-&-&0.18&0.18\\ 
                                         &$A^{D^+_s\rightarrow\phi\pi^+}_{\rm rec}$ binning&-&-&0.03&0.03\\ 
                                      &$M(K^+K^-)$ window &-&-&0.03&0.03\\ 
\raisebox{1.6ex}[0cm][0cm]{$A^{K^-}_{\epsilon}$} &$D^0\rightarrow K^-\pi^+$ statistics     &-&-&0.06&0.06\\ 
                                         &$A^{K^-}_{\epsilon}$ binning &-&-&0.04&0.04\\ 
                                         &Possible $A_{CP}^{D^0\rightarrow K^-\pi^+}$&-&-&0.01&0.01\\ \hline
$\cos\theta^{\rm CMS}_{D^+_{(s)}}$ binning&&-&-&0.06&0.06\\ 
Fitting                                  &&0.04&0.27&0.12&0.05\\ 
$K^0/\bar{K}^0$-material effects    &&0.06&0.06&0.06&0.06\\ \hline
Total                                    &&0.20&0.33&0.25&0.22\\ 
\end{tabular}     

\end{ruledtabular}
\end{table*}

The dominant source of systematic uncertainty in the
$A_{CP}^{D^+_{(s)}\rightarrow K^0_S K^+}$ measurement is the uncertainty in
$A^{K^-}_{\epsilon}$, which has several sources: the systematic uncertainty in
the $A_{\rm rec}^{D^+_s\rightarrow\phi\pi^+}$ measurement (0.18\%); statistics
of the $D^0\rightarrow K^-\pi^+$ sample (0.06\%); the systematic uncertainty
due to the choice of binning for the 2-D map of $A^{K^-}_{\epsilon}$ (0.04\%);
and a possible $A_{CP}$ in the $D^0\rightarrow K^-\pi^+$ final state from the
interference between decays with and without $D^0-\bar{D}^0$ mixing. The latter
uncertainty is estimated from the 95\% confidence level upper limit on the $CP$
violating asymmetry, $A_{CP}=-y\sin\delta\sin\phi\sqrt{R}$~\cite{PETROV}, using
the world average of $D^0-\bar{D}^0$ mixing and $CP$ violation
parameters~\cite{HFAG} and is found to be 0.01\%. We also consider different
$\cos\theta^{\rm CMS}_{D^+_{(s)}}$ binnings to estimate the systematic
uncertainty due to the choice of $\cos\theta^{\rm CMS}_{D^+_{(s)}}$ binning
(0.06\%). Systematic uncertainties due to the fitting procedure and
$K^0/\bar{K}^0$-material effects are described above and included in
Table~\ref{TABLE:SYSTOTAL}, where the total systematic uncertainties of the
$A_{CP}$ measurements are summarized. Combining all systematic uncertainties in
quadrature, we obtain $A_{CP}^{D^+\rightarrow K^0_S
  K^+}=(-0.16\pm0.58\pm0.25)\%$ and $A_{CP}^{D^+_s\rightarrow K^0_S
  K^+}=(+0.12\pm0.36\pm0.22)\%$ where the first uncertainties are statistical
and the second are systematic. Table~\ref{TABLE:SUMMARY} summarizes our
results, present best measurements~\cite{CLEOC}, and expected $A_{CP}$ from
$K^0-\bar{K}^0$ mixing~\cite{PDG2008}.
\begin{table}[htbp]
\caption{Summary of the $A_{CP}$ measurements. The first uncertainties in the
  second and third columns are statistical and the second are systematic. DCS
  decay contributions are ignored for the decays denoted by $\dagger$'s in the
  fourth column.}
\label{TABLE:SUMMARY}
\begin{ruledtabular}
\begin{tabular}{cccc} 

                           &Belle (\%)&Ref.~\cite{CLEOC} (\%) &Ref.~\cite{PDG2008} (\%) \\ \hline
$A^{D^+\rightarrow K^0_S\pi^+}_{CP}$  &$-0.71$$\pm$$0.19$$\pm$$0.20$&~\,$-1.3$$\pm$$0.7$$\pm$$0.3$&~$-0.332^\dagger$\\ 
$A^{D^+_s\rightarrow K^0_S\pi^+}_{CP}$&$+5.45$$\pm$$2.50$$\pm$$0.33$&$+16.3$$\pm$$7.3$$\pm$$0.3$  &$+0.332$\\ 
$A^{D^+\rightarrow K^0_S K^+}_{CP}$   &$-0.16$$\pm$$0.58$$\pm$$0.25$&~\,$-0.2$$\pm$$1.5$$\pm$$0.9$&$-0.332$\\ 
$A^{D^+_s\rightarrow K^0_S K^+}_{CP}$ &$+0.12$$\pm$$0.36$$\pm$$0.22$&~\,$+4.7$$\pm$$1.8$$\pm$$0.9$&~$-0.332^\dagger$

\end{tabular}     
\end{ruledtabular}
\end{table}

In summary, with a 673 fb$^{-1}$ data sample collected with the Belle detector
at the KEKB asymmetric-energy $e^+e^-$ collider, we have searched for $CP$
violation in the charged charmed meson decays $D^{+}_{(s)}\rightarrow
K^0_S\pi^+$ and $D^{+}_{(s)}\rightarrow K^0_S K^+$. No evidence for $CP$
violation is observed. Our results are consistent with the SM (see
Table~\ref{TABLE:SUMMARY}) and provide the most stringent constraints to date
on models for beyond the SM $CP$ violation in these decays~\cite{BIGI}.

We thank the KEKB group for excellent operation of the accelerator, the KEK
cryogenics group for efficient solenoid operations, and the KEK computer group
and the NII for valuable computing and SINET3 network support. We acknowledge
support from MEXT, JSPS and Nagoya's TLPRC (Japan); ARC and DIISR (Australia);
NSFC (China); DST (India); MEST, KOSEF, KRF (Korea); MNiSW (Poland); MES and
RFAAE (Russia); ARRS (Slovenia); SNSF (Switzerland); NSC and MOE (Taiwan); and
DOE (USA).


\begin{thebibliography}{99}
\bibitem{CC}
Throughout this Letter the charge conjugate decay mode is also implied unless
stated otherwise.
\end{thebibliography}
\end{document}